\documentclass[preprint,12pt]{elsarticle}
\usepackage{graphicx}
\usepackage{color}
\usepackage[mediumqspace,amssymb]{SIunits}
\usepackage{dcolumn}

\def\etal{\textit{et al. }}

\def\asq#1{ {\color{red} {..(??. #1 .??)..} } }         
\def\asq2#1{ {\color{red} {? #1 ?} } }         

\journal{International Journal of Mass Spectrometry}

\begin{document}

%
%

\begin{frontmatter}

\title{Modified binary encounter Bethe model for electron-impact ionization}

\author{M. Guerra$^1$}
\ead{mguerra@campus.fct.unl.pt} 

\author{F. Parente$^1$}
\ead{facp@fct.unl.pt} 

\author{P. Indelicato$^2$}
\ead{paul.indelicato@spectro.jussieu.fr} 

\author{J. P. Santos$^1$}
\ead{jps@fct.unl.pt}

\address{$^1$ Centro de F\'isica At\'omica, CFA, Departamento de   F\'isica, Faculdade de Ci\^encias e Tecnologia, FCT, Universidade Nova de Lisboa, 2829-516 Caparica, Portugal}
\address{$^2$   Laboratoire Kastler Brossel, \'Ecole Normale Sup\'erieure, CNRS, Universit\'e P. et M. Curie -- Paris 6, Case 74; 4, place Jussieu, 75252 Paris CEDEX 05, France }


\begin{abstract}

Theoretical expressions  for ionization cross sections by electron impact based on the binary encounter Bethe (BEB) model, valid from ionization threshold up to relativistic energies, are proposed. 

The new modified BEB (MBEB) and its relativistic counterpart (MRBEB) expressions are simpler than the BEB (nonrelativistic and relativistic) expressions because they require only one atomic parameter, namely the binding energy of the electrons to be ionized, and use only one scaling term for the ionization of all sub-shells.

The new models are used to calculate the K-, L- and M-shell ionization cross sections by electron impact for several atoms with $Z$ from 6 to 83.
Comparisons with all, to the best of our knowledge, available experimental data show that this model is as good or better than other models, with less complexity.

 \end{abstract}
 

\end{frontmatter}





  
\section{Introduction}

Knowledge of ionization and excitation cross sections is of fundamental importance for understanding collision-dynamics and electron-atom interactions, as well as in several applied fields such as radiation science, plasma physics, astrophysics and also elemental analysis using X-ray fluorescence (XRF), Auger electron spectroscopy (AES), electron energy loss spectroscopy (EELS) and electron probe microanalysis (EPMA). These areas of study need enormous and continuous quantities of data, within a certain accuracy level, for different targets over a wide range of energy values.

Electron impact ionization and excitation have been actively studied by many research groups since the 1920's. Most of the work produced was based on classical collision theory, and several first principle theories were developed \cite{14, 86, 48, 6, 8, 15}. The most important work in the field of electron-atom collision was made by Bethe (1930) who derived the correct form of the ionization cross section shape for  high-energy collisions \cite{6} using the plane-wave Born approximation (PWBA). Since then, several empirical and semi-empirical models have been proposed to describe electron impact ionization of atoms and molecules \cite{16, 17, 18, 19, 20, 47}, and several reviews on them were published \cite{21, 9}. However, each of these models works only on a limited range of target atoms and/or electron energy values and accuracies are in most cases very low. With the advance of quantum mechanical computational methods, some very accurate \textit{ab initio} calculations were performed. Nevertheless, these calculations are very time-consuming, limiting the domain of applicability of such models \cite{23, 24, 25, 26}.

In the last years, many analytical formulas have been developed to overcome these difficulties, some of them empirical  \cite{27, 28, 29} and others derived from first principles \cite{10, 11, 118, 32}.

The binary-encounter-Bethe (BEB) model proposed by Kim and Rudd~\cite{10} successfully combines the binary-encounter theory with the dipole interaction of the Bethe theory for fast incident electrons~\cite{6}, and meets the above mentioned requirements.
The BEB method, using an analytic formula that requires only the incident particle energy ($T$), the target particle's binding energy ($B$) and the target particle's kinetic energy ($U$), generates direct ionization cross section curves for neutral atoms, which are reliable in intensity ($\pm$ 20\%) and shape from the ionization threshold to a few keV in the incident energy \cite{152, 151}, or to thousands keV~\cite{35} if we consider its relativistic version (RBEB)~\cite{11}.

The factor $1/(T+U+B)$ was the only \textit{ad-hoc} term considered in the BEB model (cf. Eq. (57) in Ref. ~\cite{10} ), accounting for the projectile's kinetic energy change upon entering the atomic cloud.


Although this type of scaling has been inserted in several theories such as the PWBA \cite{12}, its success remains to be explained, even though it is an practical way to account for the electron exchange, distortion and polarization effects that are absent in the first-order PWBA.

Kim and Rudd \cite{10} noticed that they had to modify the scaling of the BEB/RBEB models. 
Comparisons to experimental data \cite{35} suggested that a simple average of the BEB cross sections with the $1/(T+U+B)$  and $1/T$ terms reproduces the experimental K-shell ionization cross section data at low to intermediate $Z$ values, and the results obtained with the classical term $1/T$ follow closely the experimental data for  L-Shell ionization. 
Thus, in order to take advantage of the success of the BEB/RBEB models, it is necessary to choose one of the terms $1/T$, $1/(T+U+B)$ and $1/2[1/T+1/(T+U+B)]$ according to the sub-shell to be ionized.

In this work, we use a different scaling for the BEB/RBEB models, in which, instead of using several scaling terms depending the ionization sub-shell, we adopt a  $1/(T+C)$ term for all sub-shells, where $C$ is a constant for each $Z$.
This constant is related with the energy change of the incident electron in the field of the nucleus and the bound electrons of the target atom. 

This article is organized as follows.  A brief outline of the underlying theory is presented in Sec.~\ref{theory}. The results are compared with available experimental and theoretical data in Sec.~\ref{results}. The conclusions are presented in Sec.~\ref{conclusions}.


\section{Theory}
\label{theory}

The relativistic theory of the BEB and RBEB models is given in detail in Refs. ~\cite{10,11}.
Below, therefore, we restrict ourselves to a rather brief account of the basic expressions, just enough for discussing the role of the scaling denominator in the ionization cross sections computation.

The term $1/T$ in the Bethe cross section was included originally to normalize the cross section to the incoming electron flux per unit area perpendicular to the incident beam direction. 

This term was modified by Burguess~\cite{90,87}, and later by Vriens~\cite{87, 90,  88, 33}, who replaced it by $1/(T+U+B)$, with the argument that the effective kinetic energy of the incident electron seen by the target is $T$ plus the energy of the bound electron. This denominator can be seen as the scaling factor to represent the correlation between the two colliding electrons. 
Although the BEB and RBEB models have been very successful in reproducing the ionization cross sections, as mentioned previously, the scaling factor may be adapted in order to take into account where ionization takes place.

In the model presented in this article we replace all the used scaling factors in the BEB/RBEB models by the $1/(T+C)$, where $C$ is a factor that depends only on $Z$.

Considering that the $C(Z)$ function in the term $1/[T+C(Z)]$  is related to the shielding of the nuclear charge by the bound electrons of the target atom, and that the binding energy of the K-shell electrons in neutral atoms (in a.u.)  scales as $0.4240 Z^{2.1822}$ (Casnati \etal \cite{148}), we may assume that  $C(Z)$  should have an almost quadratic form.
Therefore, as a first approximation, we adopt  $C(Z)$ to be equal to the hydrogenic energy levels expression, i.e.,  $C(Z)=Z_{\textrm{eff}}^2/(2n^2)$, where  $n$ the principal quantum number, and $Z_{\textrm{eff}}$ is the effective nuclear charge that accounts for the electronic shielding and electronic correlation.

Moreover, in order to emulate the energy change of the incident electron when it penetrates the electronic cloud, we assume a linear combination of the corresponding sub-shell hydrogen-like energy levels for the function $C(Z)$, which, in atomic units, can be written as 
\begin{equation}
C_{n\ell j}(Z)=a \frac{Z_{\textrm{eff,}n\ell j}^{2}}{2 n^2} + b \frac{Z_{\textrm{eff,}n' \ell' j'}^{2}}{2 n'^2},  
\label{nnline_int}
\end{equation}
where $a$ and $b$ are constants.
An analysis of the experimental results across the whole $Z$ spectra leads to the use of $a=0.3$ and $b=0.7$.

From the data published by Clementi \etal\cite{39, 38}, we have obtained $Z_{\textrm{eff},1s}=0.9834 Z - 0.1947$ and $Z_{\textrm{eff},2s}=0.7558 Z - 1.1724$. Replacing these functions in Eq. (\ref{nnline_int}), we get for K-shell ionization
\begin{equation}
C_{1s1/2}(Z) = 0.126 - 0.213\ Z + 0.195\ Z^2 .
\label{1s2s}
\end{equation}

In the cases where the $Z_{\textrm{eff}}$ is not known, we may use the well-known approximation that considers the effective nuclear charge to be given by the atomic number minus the inner electrons up to the sub-shell being ionized.

 %
%


\subsection{Modified binary encounter Bethe model}

The modified binary encounter Bethe model (MBEB) total ionization cross section, in reduced units, is written as
\begin{equation}
\sigma _{\textrm{MBEB}}=\frac{S}{t+c}\left[ \frac{1}{2}\left( 1-\frac{1}{t^{2}} \right) \ln t+\left( 1-\frac{1}{t}\right) -\frac{\ln t}{t+1}\right] ,
\label{MBEB}
\end{equation}
where the reduced units are expressed as
\begin{eqnarray}
t &=&T/B,  \nonumber \\
c &=&(C/B)2R,  \nonumber \\
S &=&4\pi a_{0}^{2}N(R/B)^{2}.  
\label{reducedunits} 
\end{eqnarray}
In Eq. (\ref{reducedunits}), $C$ is the scaling constant given by Eq. (\ref{nnline_int}), $N$ is the occupation number, $a_{0}$ is the Bohr's radius ($5.29\times 10^{-11}$ m), and $R$ is the Rydberg energy ($13.6$ eV).

The relativistic counterpart of the modified binary encounter Bethe model (MRBEB) reads
\begin{eqnarray}
\sigma _{\textrm{MRBEB}} &=&\frac{4\pi a_{0}^{2}\alpha ^{4}N}{\left( \beta
_{t}^{2}+c\beta _{b}^{2}\right) 2b^{\prime }}\left\{ \frac{1}{2}\left[ \ln
\left( \frac{\beta _{t}^{2}}{1-\beta _{t}^{2}}\right) -\beta _{t}^{2}-\ln
\left( 2b^{\prime }\right) \right] \left( 1-\frac{1}{t^{2}}\right) \right.
\nonumber \\
&&\left. +1-\frac{1}{t}-\frac{\ln t}{t+1}\frac{1+2t^{\prime }}{\left(
1+t^{\prime }/2\right) ^{2}}+\frac{b^{\prime 2}}{\left( 1+t^{\prime
}/2\right) ^{2}}\frac{t-1}{2}\right\} ,  
\label{MRBEB}
\end{eqnarray}
where
\begin{eqnarray}
\beta _{t}^{2} &=&1-\frac{1}{\left( 1+t^{\prime }\right) ^{2}}, \ \ \ \ \ \ \
\ \ \ \ \ \ t^{\prime }=T/mc^{2},   \nonumber\\
\beta _{b}^{2} &=&1-\frac{1}{\left( 1+b^{\prime }\right) ^{2}}, \ \ \ \ \ \ \
\ \ \ \ \ \ b^{\prime }=B/mc^{2},  
\label{units}
\end{eqnarray}
and $\alpha $ is the fine structure constant, $c$ is the speed of light in vacuum, and $m$ is the electron mass.

 
\section{Results}
\label{results}

The present MBEB/MRBEB models produce reliable cross sections between the threshold and the peak without using any experiment-dependent parameters.

As an illustration, we apply the nonrelativistic MBEB and relativistic MRBEB expressions  to the K-shell ionization of C, Ne, Si, Sc, Ti, V, Cr, Fe, Zn, Co, Sr, and Ag, to the L-shell ionization of Se, Kr, Ag, Sb, Xe, and Ba, and to the M-shell ionization of Pb and Bi.

Contrary to the BEB/RBEB models, which require two input parameters ($B$ and $U$), the MBEB/MRBEB models require only the knowledge of one parameter, the binding energy $B$.  For the binding energies of inner-shell electrons, one can use experimental values \cite{84} to match experimental thresholds precisely, or theoretical binding energies from Dirac-Fock wave functions that are reliable to 1\% or better in general. 
The values of $B$ of the elements studied in this work are listed in Table \ref{tab_02}.  
For the carbon atom the K-shell binding energy was taken from Ref. \cite{98}, while the remaining elements K-shell binding energies were obtained from Ref. \cite{84}. 
The L- and M-shell binding energies were evaluated using the MDFGME code developed by J. P. Desclaux and P. Indelicato \cite{92,93}.

The electron occupation number was set to $N=2$ for $s_{1/2}$ and $p_{1/2}$ orbitals, $N=4$ for $p_{3/2}$ and  $d_{3/2}$ orbitals and $N=6$ for $d_{5/2}$ orbitals. 

 
\subsection{K-shell ionization}
\label{results_k}

\begin{figure}[ht]
	\centering
		\includegraphics[width=1\textwidth]{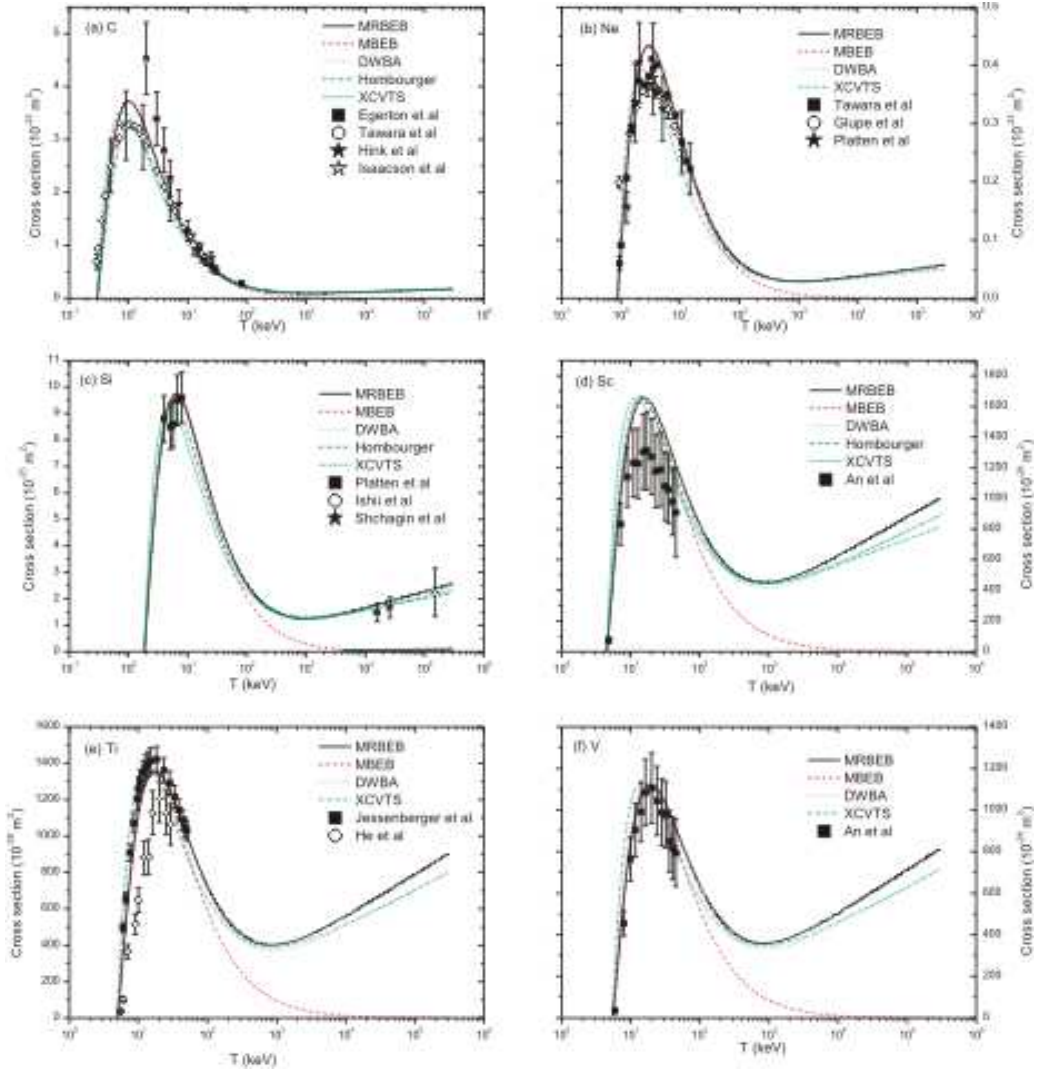}
	\caption{Electron impact K-shell ionization cross sections for (a) C, (b) Ne, 
	(c) Si, (d) Sc, (e) Ti, (f) V. 
	Thick solid curve, present MRBEB cross section Eq. (\ref{MRBEB}); 
	dash-dash curve, MBEB cross section Eq. (\ref{MBEB})
	dot-dot curve, DWBA by Bote \etal\cite{28};
	dot-dash curve, relativistic empirical formula by Hombourger \cite{91}; 
	short dot-dash curve, XCVTS  semiempirical formula by Haque \etal\cite{29};
	Experimental data by 
	Egerton \etal\cite{46}, 
	Tawara \etal\cite{4}, 
	Hink \etal\cite{49}, 
	Isaacson \etal\cite{45}, 
	Glupe \etal\cite{1}, 
	Platten \etal\cite{3}, 
	Ishii \etal\cite{55}, 
	Kamiya \etal\cite{54}, 
	Hoffman \etal\cite{52},
	Shchagin \etal\cite{56},
	He  \etal\cite{60}, 
	Jessenberger  \etal\cite{59}, 
	and
	An  \etal\cite{61}.}
	\label{fig:NewFig1MRBEB}
\end{figure}

\begin{figure}[ht]
	\centering
		\includegraphics[width=1\textwidth]{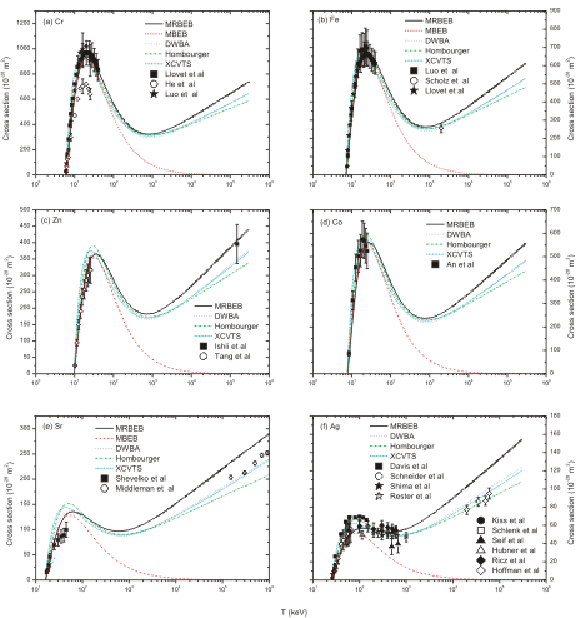}
	\caption{Electron impact K-shell ionization cross sections for (a) Cr, (b) Fe, (c) Zn, (d) Co, (e) Sr and (f) Ag.
  Thick solid curve, present MRBEB cross section Eq. (\ref{MRBEB}); 
	dash-dash curve, MBEB cross section Eq. (\ref{MBEB})
	dot-dot curve, DWBA by Bote \etal\cite{28};
	dot-dash curve, relativistic empirical formula by Hombourger \cite{91}; 
	short dot-dash curve, XCVTS semiempirical formula by Haque \etal\cite{29};
	Experimental data by
	Llovet  \etal\cite{63},
	He  \etal\cite{60},
	Luo  \etal\cite{65} (Cr),
	Luo  \etal\cite{62} (Fe),
	Scholz  \etal\cite{64},
	Ishii \etal\cite{55},
	Tang \etal\cite{78},
	An  \etal\cite{69}, 
	Shevelko \etal\cite{80}, 
	Middleman \etal\cite{70},
	Davis  \etal\cite{7}
	Schneider \etal\cite{75},
	Shima \etal\cite{72},
	Rester \etal\cite{74},
	Kiss \etal\cite{79}, 
	Schlenk \etal\cite{81}, 
	El Nasr \etal\cite{66}, 
	Hubner \etal\cite{77},
	Ricz \etal\cite{82}, 
	and 
	Hoffman \etal\cite{52}.}
	\label{fig:NewFig2MRBEB}
\end{figure}

On Fig. \ref{fig:NewFig1MRBEB} (for C, Ne, Si, Sc, Ti, and V) and Fig. \ref{fig:NewFig2MRBEB} ( for Cr, Fe, Zn, Co, Sr and Ag), we compare the present MBEB [Eq. (\ref{MBEB})] and MRBEB cross sections [Eq. (\ref{MRBEB})]  to all available experimental data, to the empirical cross sections by 
Hombourger \etal\cite{91}, Haque \etal\cite{29}, and to the analytical model by Bote \etal\cite{28}, which results from a fit to a database of cross sections calculated using the plane-wave (PWBA) and distorted-wave (DWBA) Born approximations.
For overvoltages ($t=T/B$) lower than 16, the fit was done to the DWBA database, and for $t>16$  the PWBA database was used, since, for high-energies, the difference between the DWBA and PWBA cross sections is negligible. 
The DWBA/PWBA model, labeled as DWBA for simplicity,  provides ionization cross section values that agree with those in the DWBA/PWBA database to within about 1\%, except for projectiles with near-threshold energies.
Since both the Hombourger \etal model and the XCVTS model of Haque \etal are empirical, the range of validity of such models is limited by the availability of experimental data.
Furthermore, the XCVTS model uses a scaling term with different coefficients for different shells as in the unmodified BEB/RBEB expressions.

In the analysis of Figs. \ref{fig:NewFig1MRBEB} and \ref{fig:NewFig2MRBEB}, as discussed previously by Santos \etal \cite{35}, caution is warranted when comparing the experimental and theoretical data represented. 
Experimental data are mainly obtained through the detection of X-rays or Auger electrons emitted when bound electrons fill the K-shell vacancies created by electron impact. However, K-shell vacancies can be created not only by direct ionization but also by excitations of K electrons to unoccupied bound states.  Since most theories, including the MBEB/MRBEB models, are designed for only direct ionization by electron impact, experimental data may exceed the theoretical data by the amount due to excitations of K electrons to bound levels.  Therefore, unless experimental data have explicitly excluded the K-shell vacancies created by excitation, comparisons of theories and experiments may have an inherent ambiguity of $\sim$10\%.

Below we discus the  cases that we analyzed. In order to compare the experimental values to the different theoretical results, we used the reduced $\chi ^2$, $Q$, defined by $Q=\chi ^2/ \nu$, where $ \nu$ is the number of experimental data points:
\begin{itemize}
\item Carbon: The relativistic and nonrelativistic cross sections are almost identical for $T<1$ keV. The present MRBEB cross section, the DWBA and the XCTVS results are in good agreement with the experimental data by Egerton \etal\cite{46}, Tawara \etal\cite{4}, and Isaacson \etal\cite{45} (with the reduced $\chi ^2$, $Q$, equal to 0.91, 0.65 and 0.73, respectively), while the experimental data by Hink \etal\cite{49} display an increasing trend toward lower $T$ not seen in any other theory or experiment.
\item Neon: The relativistic and nonrelativistic cross sections are almost identical for $T <$ 100 keV. 
The theoretical cross sections are in fairly good agreement with  experimental data by Tawara \etal\cite{4}, Glupe \etal\cite{1}, and Platten \etal\cite{3}. 
\item Silicon: We see the beginning of the relativistic rise at $T >$ 100 keV, which is not followed by the nonrelativistic MBEB. In this high $T$ region, all theoretical relativistic data agree with the experimental data by Ishii \etal\cite{55} and Shchagin \etal\cite{56}, with $Q$ values from 0.4 (Hombourger) to 0.7 (DWBA).
\item Scandium: The experimental results by An \etal\cite{61} are not in agreement with any of the theories presented here, so new experimental data are required to better understand this case.
\item Titanium: The experiments are divided into two groups. The experimental cross sections by Jessenberger  \etal\cite{59} lie above all theoretical data in the peak region, while the experimental cross sections by He  \etal\cite{60} are lower than all theoretical data. 
\item  Vanadium: The MRBEB cross section values for vanadium are in good agreement with the experimental data  by An  \etal\cite{61}, having the lowest $Q$ value of all theoretical models, which ranges from 20.9 to  130.5.
\item  Chromium: We notice that all experimental data except the  one from He  \etal\cite{60} for chromium agree with the represented theoretical models, confirming the trend of the experimental data by He  \etal observed in Ti. 
\item Iron: Although there is a general agreement between the theoretical data and the experimental results, the MRBEB model underestimates slightly the ionization cross sections in the peak region.
\item Zinc: The MRBEB cross sections are in good agreement with the experimental data by Tang \etal\cite{78} at low $T$, and with the only experimental value at high $T$ from Ishii \etal\cite{55}. This is confirmed by the low $Q$ value of 1.1 that we find, to be compared to the high value of 18.3 for the XCVTS model. Nevertheless, the MRBEB values become larger than the other three theoretical cross section values beyond $T=1$ MeV. There is thus a strong need of new experiment  for $T>1$ MeV is desirable to distinguish different predictions from different theories.
\item Cobalt: The experimental data by An \etal\cite{69} agree very well with the MRBEB model, from threshold to the ionization peak, which produces the lowest $Q$ value of all models in a range from 0.2 to 10.6.
\item Strontium: The theoretical data disagree among them and with the experimental data. However, we observe that the MRBEB model ($Q$=3.2) follows more closely the experimental data by Shevelko \etal\cite{80} at low $T$, while the DWBA ($Q$=6.3) and the Hombourger ($Q$=10.0) models follow  more closely the experimental data by Middleman \etal\cite{70} at high $T$. 
\item Silver: Ten sets of experimental data  are compared with the MRBEB cross sections and other theories. Again, experiments are divided into groups near the peak. The experimental data by Davis  \etal\cite{7} agree well with the Hombourger cross sections. The experimental data by Schneider \etal\cite{75}, Kiss \etal\cite{79}, Hoffman \etal\cite{52} agree with the MRBEB cross sections. The data by el Nasr \etal \cite{66} and Hubner \etal\cite{77} disagree with all the presented theoretical cross sections.
Although all theoretical cross sections agree in the vicinity of $T=500$ keV, the difference between the present MRBEB cross section values and the other theoretical relativistic  cross section values  is widening at $T=1$ MeV, amplifying the trend observed in Zn and Sr. The silver atom is another example for which definitive measurements would help to distinguish different theories.
\end{itemize}
%


\subsection{L- and M-shell ionization}
\label{results_LM}

In order to investigate the range of applicability of the approach presented in this work besides the K-shell ionization, we have also applied the MBEB and MRBEB models to the L-shell ionization of Se, Kr, Ag, Sb, Xe and Ba, and to M-shell ionization of Pb and Bi.

\begin{figure}[ht]
	\centering
		\includegraphics[width=1\textwidth]{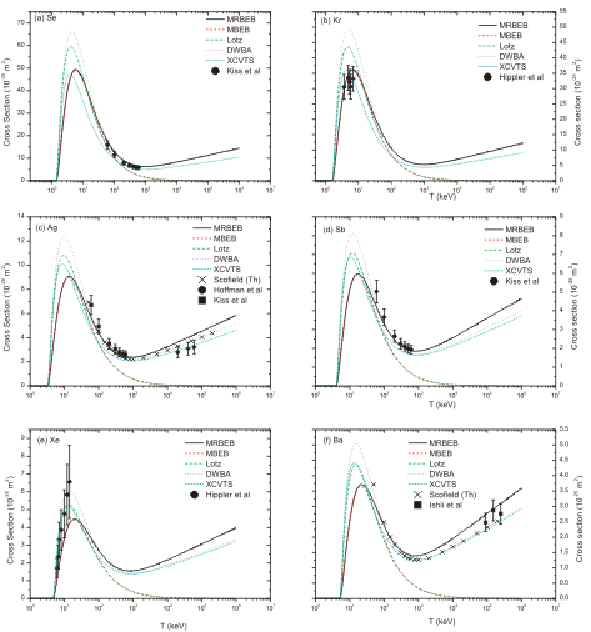}
	\caption{Electron impact L-shell ionization cross sections for (a) Se, (b) Kr, (c) Ag, (d) Sb, (e) Xe and (f) Ba.
  Thick solid curve, present MRBEB cross section Eq. (\ref{MRBEB}); 
	dash-dash curve, MBEB cross section Eq. (\ref{MBEB})
	dot-dash curve, relativistic empirical formula by Lotz \cite{47,153}; 
	dot-dot curve, DWBA by Bote \etal\cite{28};
	short dot-dash curve, XCVTS semiempirical formula by Haque \etal\cite{29};
	$\times$, DWBA values by Scofield \etal\cite{150};
	Experimental data by
	Ishii \etal\cite{55},
	Kiss \etal\cite{79},
	Hippler  \etal\cite{131},
	 and
	Hoffman \etal\cite{2}.}
	\label{fig:NewFig3MRBEB}
\end{figure}

\begin{figure}[ht]
	\centering
		\includegraphics[width=1 \textwidth]{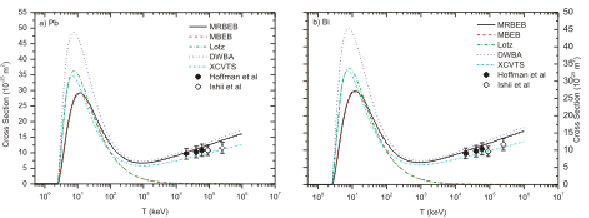}
	\caption{Electron impact M-shell ionization cross sections for (a) Pb and (f) Bi.
  Thick solid curve, present MRBEB cross section Eq. (\ref{MRBEB}); 
	dash-dash curve, MBEB cross section Eq. (\ref{MBEB})
	dot-dash curve, relativistic empirical formula by Lotz \cite{47,153}; 
	dot-dot curve, DWBA by Bote \etal\cite{28};
	short dot-dash curve, XCVTS semiempirical formula by Haque \etal\cite{29};
	Experimental data by
	Ishii  \etal\cite{55},
	and 
	Hoffman \etal\cite{2}
	.}
	\label{fig:NewFig4MRBEB}
\end{figure}

On Fig. \ref{fig:NewFig3MRBEB} and  Fig. \ref{fig:NewFig4MRBEB} the MBEB and MRBEB cross sections for the L- shell (for Se, Kr, Ag, Sb, Xe, and Ba) and M-shell (for Pb and Bi), respectively, are displayed as well as the theoretical results obtained with the DWBA, XCVTS and Lotz~\cite{47,153} models, and by Scofield~\cite{150}, and the experimental available data for the analyzed elements. 
The Lotz empirical expression, proposed more than 30 years ago, is one of the most successful formulas for calculating total direct ionization of any given state. 

Concerning the L-Shell ionization, we notice that the MRBEB cross sections are in good agreement with the experimental data for the analyzed elements, except Xe, having the lowest $Q$ value for Se, Kr, Sb and Ba (1.2, 0.6, 0.5, 0.6, respectively), and the second lowest for Ag (1.4). The theoretical data disagree among them, namely in the peak region; the DWBA values produce the highest peak, followed in equal ground by the XCVTS and the Lotz curves, and finally by the MBEB and MRBEB curves.
It should be pointed out that the experimental data by Hippler \etal~\cite{131} for Xe exhibits the greater uncertainty (about 30\%) among the studied cases. This uncertainty is less than 17\% for the other elements.
The experimental data for the M-shell ionization is scarce and exist only for high incident electron energies, in the relativistic regime ($T>10^4$ keV). 
In this high region, all theoretical relativistic data agree with the experimental data by Ishii \etal\cite{55} and Hoffman \etal~\cite{52}, with $Q$ values equal to 0.4 (XCVTs), 0.7 (MRBEB), and 1.7 (DWBA) for Pb, and  0.3 (XCVTs), 0.9 (MRBEB), and 2.0 (DWBA) for Bi. 
The comparison among the theoretical data have the same outcome obtained for the L-Shell.

%


\section{Conclusions}
\label{conclusions}

The new MBEB and MRBEB models presented in this work require only one atomic parameter, namely the binding energy of the electrons to be ionized, and, contrary to the BEB/RBEB models, use only one scaling term (1/(T + C)) for the ionization of all sub-shells.

The MBEB and MRBEB expressions were used to obtain the K-, L-, and M shell ionization cross sections by electron impact for  several atoms with Z from 6 to 83.
 
We  pointed out that the comparison of the MRBEB cross sections  to experimental values contains inherent ambiguities, because the MRBEB model predicts cross sections for the direct ionization of electrons of a definite sub-shell, 
while most experimental data are based on all sub-shell vacancies created by direct ionization as well as excitations to bound levels.

As  show on Figs. \ref{fig:NewFig1MRBEB}, \ref{fig:NewFig2MRBEB}, \ref{fig:NewFig3MRBEB}, and \ref{fig:NewFig4MRBEB}, relativistic effects become increasingly important as the binding energies of the elements increase. Hence, relativistic theory must be used for treating both atomic structure and collision dynamics for medium to heavy atoms.

The presented comparisons  show that the MRBEB model produces reliable K-, L- and M-shell ionization cross sections between the threshold and several MeV  with an accuracy of $\sim$20\%, or better, without using empirical parameters.

The simple relativistic MRBEB expression presented in this article provides a continuous coverage of K-, L- and M-shell ionization cross sections by electron impact from the threshold to relativistic incident energies, making this expression ideally suited for modeling systems  where ionization cross sections for a wide range of incident energies are required, such as fusion plasmas.

%
%
%
\section*{Acknowledgements}

This research was supported in part by FCT (PEst-OE/FIS/UI0303/2011, Centro de F'sica At\'omica), by the French-Portuguese collaboration (PESSOA Program, contract no 441.00), by the Ac\c{c}\~oes Integradas Luso-Francesas (contract no F-11/09), and by the Programme Hubert Curien.
M.G. acknowledges the support of the FCT, under Contract No. SFRH/BD/38691/2007.
Laboratoire  Kastler Brossel is ``Unit\'e Mixte de Recherche du CNRS, de l' ENS et de l'UPMC n$^{\circ}$ 8552''. 
%
%

\section*{References}

\bibliography{MGuerra}{}
\bibliographystyle{elsarticle-harv}

%
\newpage
\begin{table}
\caption{
Binding energy $B$  values for the K-, L- and M-shells. The $B$ value for C is from Ref. \cite{98}. The remaining K-shell and L-shell $B$ values are from Ref. \protect\cite{84}. The M-shell binding energies were evaluated using the MDFGME code ~\cite{92,93}  
\protect\label{tab_02}}

%
%
\scalebox{0.7}{
\begin{tabular}{cccccccccc}
Element & & & & \multicolumn{3}{c}{$B$(eV)}  \\
\cline{2-10} 
  &  K-Shell & \multicolumn{3}{c}{L-shell} &  \multicolumn{5}{c}{M-shell}  \\
  \cline{3-5} \cline{6-10} 
 & & L1 & L2 & L3 & M1 & M2 & M3 & M4 & M5 \\  
C               &       296.07   &	&	&	&	&	&	&	& \\
Ne              &       866.90   &	&	&	&	&	&	&	& \\
Si              &       1840.05  &	&	&	&	&	&	&	& \\
Sc              &       4489.37  &	&	&	&	&	&	&	& \\
Ti              &       4964.58  &	&	&	&	&	&	&	& \\
V               &       5463.76  &	&	&	&	&	&	&	& \\
Cr              &       5989.02  &	&	&	&	&	&	&	& \\
Fe              &       7110.75  &	&	&	&	&	&	&	& \\
Co              &       7708.75  &	&	&	&	&	&	&	& \\
Zn              &       9660.76  &	&	&	&	&	&	&	& \\
Se							&								 & 1652.44 & 1474.72 & 1433.98 &	&	&	&	&  \\
Kr							&								 & 1916.30 & 1729.66 & 1677.25 &	&	&	&	&  \\
Sr              &       16107.20 &	&	&	&	&	&	&	& \\
Ag              &       25515.59 & 3807.34 & 3525.83 & 3350.96 &	&	&	&	& \\
Sb							&								 & 4698.44 & 4381.90 & 4132.33 &	&	&	&	&  \\
Xe							&								 & 5452.89 & 5103.83 & 4782.16 &	&	&	&	&  \\
Ba							&								 & 5995.90 & 5623.29 & 5247.04 &	&	&	&	&  \\
Pb							&								 &	&	&	& 3905.53 & 3601.14 & 3110.21 & 2628.17 & 2525.49 \\
Bi							&								 &	&	&	& 4056.25 & 3744.91 & 3223.11 & 2731.84 & 2623.08 \\
\end{tabular}
}
%
\end{table}

%
%
%
\cleardoublepage
\newpage

%
%
%

%
%
%

%
%
%

%
%
%
%

%
%

\end{document}